\documentclass[twocolumn,pra]{revtex4}
\usepackage{graphicx}

\begin{document}
\title{Unambiguous pure-state identification without classical knowledge}
\author{A. Hayashi, M. Horibe, and T. Hashimoto}
\address{Department of Applied Physics\\
           University of Fukui, Fukui 910-8507, Japan}

\begin{abstract}
We study how to unambiguously identify a given quantum pure state with one of 
the two reference pure states when no classical knowledge on the reference 
states is given but a certain number of copies of each reference quantum state 
are presented. 
By unambiguous identification, we mean that we are not allowed 
to make a mistake but our measurement can produce an inconclusive result. 
Assuming the two reference states are independently distributed 
over the whole pure state space in a unitary invariant way, we determine 
the optimal mean success probability for an arbitrary number of copies of the 
reference states and a general dimension of the state space.
It is explicitly shown that the obtained optimal mean success probability 
asymptotically approaches that of the unambiguous discrimination as the number 
of the copies of the reference states increases. 
\end{abstract}

\maketitle

\newcommand{\ket}[1]{|\,#1\,\rangle}
\newcommand{\bra}[1]{\langle\,#1\,|}
\newcommand{\braket}[2]{\langle\,#1\,|\,#2\,\rangle}
\newcommand{\bold}[1]{\mbox{\boldmath $#1$}}
\newcommand{\sbold}[1]{\mbox{\boldmath ${\scriptstyle #1}$}}
\newcommand{\tr}[1]{{\rm tr}\left[#1\right]}
\newcommand{\BC}{{\bold{C}}}
\newcommand{\CS}{{\cal S}}

\section{Introduction}
In quantum mechanics one cannot perfectly clone an unknown state 
\cite{Wootters82}, which makes the problem of distinguishing quantum states 
nontrivial \cite{Helstrom76,Holevo82}.
Imagine we are presented with an unknown quantum pure state $\rho$ on a 
$d$-dimensional vector space $\BC^d$. 
Let us assume that the input state $\rho$ is guaranteed to be either one of 
two reference states $\rho_1$ and $\rho_2$, each being also a pure state on 
$\BC^d$. Then how well can we identify the input state with one of the two 
reference states?

We can consider two cases depending on what kind of information on 
the reference states is available. 
In the first case, it is assumed that we have complete classical knowledge on 
the two reference states $\rho_1$ and $\rho_2$. 
This is the standard setting of quantum-state discrimination, which was solved 
by Helstrom \cite{Helstrom76}.

On the other hand, we can also consider the case where only a certain number 
$(N)$ of copies of $\rho_1$ and $\rho_2$ are presented, with no classical 
knowledge on them available \cite{Hayashi05_identification}. 
See also related works in the case of qubits \cite{Sasaki01,Sasaki02}.
In this case, we could obtain only limited classical information on the 
reference states, since the no-cloning theorem \cite{Wootters82} does not 
allow us to increase the number of copies of the reference states.
The best we can do is to perform a positive-operator-valued measure (POVM) 
measurement on the total state  
$\rho \otimes \rho_1^{\otimes N} \otimes \rho_2^{\otimes N}$ and try to 
identify the input state $\rho$ with one of the reference states $\rho_1$ 
and $\rho_2$. If the number of copies, $N$, is infinite, the problem is 
reduced to quantum-state discrimination, since we could always obtain 
complete classical knowledge of a quantum state. 
In our previous paper \cite{Hayashi05_identification}, 
we called this problem  ``state identification" and determined the optimal 
mean identification probability for an arbitrary number ($N$) of copies of 
the reference states in a general dimension $d$.

In the standard setting of the discrimination problem, 
we are allowed to make an error and are interested in the optimal strategy 
that makes the error probability minimum. On the other hand, an error is not 
allowed in the problem of unambiguous discrimination 
\cite{Ivanovic87,Dieks88,Peres88}. 
Instead our measurement can produce one of three outputs 1, 2, or 0. 
If the output is 1(2), we are certain that the input state $\rho$ is 
$\rho_1(\rho_2)$, and the output 0 means that we do not know the identity of 
the input, which is called an inconclusive result. The optimal strategy is 
the one that minimizes the probability of the inconclusive result.  

We can also generalize the unambiguous discrimination problem to the case 
in which a finite number ($N$) of copies of the 
reference states are presented without any classical information on them, 
which will be called the problem of unambiguous state identification in this 
paper. Bergou and Hillery studied this problem in the case of qubits ($d=2$) 
when the number of copies of reference states, $N=1$ \cite{Bergou05}. 
They called the optimal strategy a programmable state discriminator since the 
strategy is not ``hard wired"  but supplied by the reference states stored in 
registers in the machine. 

The problem of the quantum-state comparison \cite{Barnett03,Chefles04} 
is related to the state identification problem. One's task here is to 
establish whether or not two quantum systems have been prepared in the 
same state. The symmetry under interchanging the systems is essential in 
the state comparison, since the combined system is symmetric when the two 
systems are in the same state, whereas it has no definite exchange symmetry 
otherwise. The exchange symmetry plays a crucial role also in the state 
identification, but in a more involved way.

In this paper we study the problem of unambiguous identification of 
pure states for an arbitrary number $N$ of copies of the reference states 
in a general dimension $d$. The two pure reference states $\rho_1$ and 
$\rho_2$ are independently distributed over the whole state space on 
$\BC^d$ in a unitary invariant way. 
The input state $\rho$ is assumed to be either one of 
the two reference states with the equal probabilities. 
We determine the optimal POVM and the 
optimal mean unambiguous identification probability, which can be explicitly 
shown to approach the mean unambiguous discrimination probability in the 
large-$N$ limit. 

\section{Mean unambiguous pure state discrimination}
In this section we average the unambiguous discrimination 
probability, assuming that the two reference states are independently 
distributed on $\BC^d$ in a unitary-invariant way. This mean unambiguous 
discrimination probability will be later compared with the mean unambiguous 
identification probability in the large-$N$ limit.

More precisely the unitary distribution of the reference states is specified 
in the following way. 
Expand a pure state as $\ket{\phi}=\sum_{i=1}^d c_i\ket{i}$ in terms of an 
orthonormal base $\{\ket{i}\}$ of $\BC^d$. The distribution is then defined to 
be the one in which the $2d$-component real vector 
$\{x_i={\rm Re}\, c_i, y_i={\rm Im}\, c_i \}$ is uniformly distributed on the 
$(2d-1)$-dimensional hypersphere of radius 1 with the integration measure 
given by 
\begin{eqnarray}
   dcdc^+ \equiv \prod_{i=1}^d
     (dx_i dy_i) \delta\left(\sum_i(x_i^2+y_i^2) -1\right).
                         \label{measure}
\end{eqnarray}
Evidently the distribution does not depend on the choice of the orthonormal 
base $\{\ket{i}\}$.

The optimal success probability of unambiguous discrimination 
of two known pure states $\rho_1=\ket{\phi_1}\bra{\phi_1}$ and 
$\rho_2=\ket{\phi_1}\bra{\phi_1}$ on $\BC^d$ is given by 
\cite{Ivanovic87,Dieks88,Peres88}
\begin{eqnarray}
  p_{\max}(\rho_1,\rho_2) = 1 - |\braket{\phi_1}{\phi_2}|.
\end{eqnarray}
We calculate the mean unambiguous discrimination probability 
\begin{eqnarray}
   p_{\max}(d) = \Big< p_{\max}(\rho_1,\rho_2) \Big>,
\end{eqnarray}
where $\langle \cdots \rangle$ means the average over $\rho_1$ and $\rho_2$, 
which are independently distributed according to the unitary distribution 
defined above.

The average $\langle |\braket{\phi_1}{\phi_2}| \rangle$ can be calculated in 
terms of an integration over a $(2d-1)$-dimensional hypersphere as 
\begin{eqnarray}
   \langle |\braket{\phi_1}{\phi_2}| \rangle = 
       \frac{ \int\! dcdc^+ \,|c_1| }{ \int\! dcdc^+ 1},
\end{eqnarray}
with the integration measure given by Eq. (\ref{measure}).

We obtain 
\begin{eqnarray}
   p_{\max}(d) = 1 - \frac{2^{d-1}(d-1)!}{(2d-1)!!}, 
                         \label{p_discrimination}
\end{eqnarray}
which is certainly less than the mean discrimination probability 
given in \cite{Hayashi05_identification}:
\begin{eqnarray}
\Big< \frac{1}{2}(1+\sqrt{1-|\braket{\phi_1}{\phi_2}|^2}) \Big> = 
                  \frac{1}{2}+\frac{d-1}{2d-1}.
\end{eqnarray}

\section{Unambiguous pure state identification}
Suppose we are given an unknown pure state $\rho$ on $\BC^d$.
We know that $\rho$ is either one of the two reference states 
$\rho_1$ and $\rho_2$ on $\BC^d$, with equal prior probabilities. 
Let us assume that we have no classical knowledge on the reference states, 
but a certain number $(N)$ of copies of each state are available.
What is the optimal strategy to unambiguously identify the 
input state with one of the reference states 
when the two reference states are independently distributed over the whole 
pure-state space in a unitary invariant way?
And what is the optimal mean probability of success?

We assume that the input state $\rho$ is prepared in system 0 and 
$N$ copies of each reference state $\rho_a$ ($a=1,2$) in systems 
$a_1,a_2,\ldots,a_N$, which will be collectively denoted by $a$. 
We specify the system which an operator acts on by the system number in the 
parentheses; namely, $\rho(0)$ means that this is an operator acting on 
system 0, for example. 

Our task is then to unambiguously distinguish two states 
$\rho_1(0) \rho_1^{\otimes N}(1) \rho_2^{\otimes N}(2)$ and 
$\rho_2(0) \rho_1^{\otimes N}(1) \rho_2^{\otimes N}(2)$.
The mean success probability of identification is given by
\begin{eqnarray}
 p^{(N)}(d) = \frac{1}{2} \sum_{a=1}^{2} \Big<
      \tr{E_a\rho_a(0)\rho_1^{\otimes N}(1)\rho_2^{\otimes N}(2) }
                          \Big>,
\end{eqnarray}
where $\{E_0,E_1,E_2\}$ is a POVM acting on the whole system and 
$\langle \cdots \rangle$ represents 
the average over $\rho_1$ and $\rho_2$ defined in the preceding section. 
When the outcome of the POVM is 
$a\ (=1,2)$, we identify the input $\rho$ with $\rho_a$ with certainty.
Outcome $0$ of the POVM means we have an inconclusive result.
Note that the POVM should be independent of $\rho_1$ and $\rho_2$ since we 
are given no classical knowledge on them.

The average over the reference states can be easily performed 
by the use of formula for the average of the $n$-fold tensor product of 
an identical pure state $\rho$
\begin{eqnarray}
   \langle \rho^{\otimes n} \rangle = \frac{\CS_n}{d_n}, \label{formula1}   
\end{eqnarray}
where $\CS_n$ is the projection operator onto the totally symmetric subspace 
and $d_n$ is its dimension, $d_n=\tr{\CS_n}={}_{n+d-1}C_{d-1}$ 
\cite{Hayashi05_estimation}. 
We find 
\begin{eqnarray}
 p^{(N)}(d) = \frac{1}{2d_{N+1}d_N} \Big(
         \tr{E_1\CS_{N+1}(01)\CS_N(2)}   
                      \nonumber \\
        +\tr{E_2\CS_N(1)\CS_{N+1}(02)}
                                    \Big),
                              \label{pS}
\end{eqnarray}
where $\CS_{N+1}(01)$ is the projector onto the totally symmetric subspace on 
systems $(0,1)=(0,1_1,1_2,\ldots,1_N)$ and other $\CS$'s are defined 
similarly. 

The POVM should satisfy the following conditions:
\begin{eqnarray}
   E_0, E_1, E_2 \ge 0,\ \ \ E_0+E_1+E_2=1
                                 \label{positivity}
\end{eqnarray}
and, for any $\rho_1$ and $\rho_2$,
\begin{eqnarray}
  \tr{E_1\rho_2(0)\rho_1^{\otimes N}(1)\rho_2^{\otimes N}(2) } &=& 0,
                                       \nonumber \\
  \tr{E_2\rho_1(0)\rho_1^{\otimes N}(1)\rho_2^{\otimes N}(2) } &=& 0,
\end{eqnarray}
which implies no error is allowed.
It is evident that the above no-error conditions are equivalent to
\begin{eqnarray}
  &&  E_1\CS_N(1)\CS_{N+1}(02) = \CS_N(1)\CS_{N+1}(02)E_1 = 0,
                                       \nonumber \\
  &&  E_2\CS_N(2)\CS_{N+1}(01) = \CS_N(2)\CS_{N+1}(01)E_2 = 0.
                                        \label{unambiguity}
\end{eqnarray}

Now we observe that the set of POVM's satisfying conditions 
(\ref{positivity}) and (\ref{unambiguity}) is convex. Namely, if 
each of two POVM's $E_a$ and $E'_a$ respects 
conditions (\ref{positivity}) and (\ref{unambiguity}), so does their 
convex linear combination $qE_a+(1-q)E'_a$ for 
any $ 0 \le q \le 1$.
And the resulting probability, Eq.(\ref{pS}), is also a convex 
combination: $p(qE+(1-q)E') = qp(E) + (1-q)p(E')$ in an obvious abbreviated 
notation. 

We exploit this convexity of the POVM to impose some symmetries on the optimal 
POVM without loss of generality. 
First we notice the problem is symmetric under the exchange between 
systems 1 and 2. Suppose a POVM $F_a$ is optimal. 
Then another POVM $F'_a$, defined by 
\begin{eqnarray}
  F'_1 = T F_2 T,\ F'_2 = T F_1 T,\ F'_0 = T F_0 T, 
\end{eqnarray}
is also legitimate and optimal. Here we introduced the exchange operator 
$T$ between systems 1 and 2.
Then a new POVM 
$ E_a = \frac{1}{2} \left( F_a + F'_a \right) $ 
is also optimal and satisfies the exchange symmetry between systems 1 and 2, 
\begin{eqnarray}
     E_2=TE_1T,\ E_0=TE_0T. \label{ex_symmetry}
\end{eqnarray}

The second symmetry we consider is the unitary symmetry of the distribution 
of the reference states. If a POVM $F_a$ is optimal, 
another POVM defined by 
\begin{eqnarray}
   U^{\otimes (2N+1)}F_a(U^+)^{\otimes (2N+1)}\ (a=0,1,2)
\end{eqnarray} 
is also legitimate and optimal for any unitary operator $U$.
Let us construct a POVM by 
\begin{eqnarray}
    E_a = \int dU U^{\otimes (2N+1)}F_a(U^+)^{\otimes (2N+1)} 
                                     \nonumber \\ 
                           (a=0,1,2),
\end{eqnarray}
where $dU$ is the normalized positive-invariant measure of the group $U(d)$.
The new POVM $E_a$ is clearly a legitimate optimal POVM.  
Furthermore, since $E_a$ commutes with  $U^{\otimes (2N+1)}$ for any $U$, 
we conclude that $E_a$ is a scalar with respect to the group $U(d)$. 
Thus we can assume that the optimal POVM satisfies the exchange symmetry of 
Eqs.~(\ref{ex_symmetry}) and is scalar with respect to the group $U(d)$.

By the exchange symmetry, the mean probability, Eq.(\ref{pS}), to be optimized 
takes the form
\begin{eqnarray}
 p^{(N)}(d) = \frac{1}{d_{N+1}d_N}\tr{E_1\CS_{N+1}(01)\CS_N(2)}.
                                       \label{pS_symmetric}
\end{eqnarray}
And the conditions $E_1$ should satisfy are given by
\begin{eqnarray}
  E_1 \ge 0,\ 1 \ge E_1+TE_1T, 
                 \label{positivity_symmetric}
\end{eqnarray}
and the no-error conditions 
\begin{eqnarray}
  E_1\CS_N(1)\CS_{N+1}(02) = \CS_N(1)\CS_{N+1}(02)E_1 = 0.
                                        \label{unambiguity_symmetric}
\end{eqnarray}
   
Finally we note that we can work in the subspace $V_{\rm sym}$, in which each 
of systems 1 and 2 is both totally symmetric. With this in mind, we set 
$\CS_N(1)=\CS_N(2)=1$ hereafter.

\section{Case of qubits ($d=2$)}
In this section we study the case of qubits ($d=2$), where the individual 
system can be regarded as a spin-$1/2$ particle and the problem reduces to the 
angular momentum recoupling. 
In the subspace $V_{\rm sym}$, each system $a\ ({=}1,2)$ consisting of $N$ 
spin-$1/2$ particles is totally symmetric, implying the total angular 
momentum of each system is $j \equiv N/2$. 

We can construct the total angular momentum of the whole $2N+1$ systems 
in two ways. First the combined system of $0$ and $1$ has the angular 
momentum $J_1=j_-\equiv j-1/2$ or $J_1=j_+\equiv j+1/2$. 
Then this intermediate angular momentum $J_1$ is coupled with the angular 
momentum $j$ of system $2$, 
resulting in the total angular momentum of the whole system $J$. 
Using the standard notation \cite{Rose57}, we write the resultant 
eigenstate with the total angular momentum $J$ and its $z$ component $M$ 
as 
\begin{eqnarray}
  \ket{A_{J_1};JM} \equiv \ket{(j\frac{1}{2})J_1,j;JM}\ 
                      \ (J_1=j_-,j_+), 
                             \label{A_base}
\end{eqnarray}
where we ordered three systems as $1 \otimes 0 \otimes 2$ on the 
right-hand side.
Note that the state $\ket{A_{j_+};JM}$ is totally symmetric in the 
subspace of systems 0 and 1 and the state $\ket{A_{j_-};JM}$ is not, 
that is, 
\begin{eqnarray}
     \CS_{N+1}(01) \ket{A_{j_+};JM} &=& \ket{A_{j_+};JM}, 
                              \nonumber \\
     \CS_{N+1}(01) \ket{A_{j_-};JM} &=& 0.
\end{eqnarray}
Another coupling scheme is that systems 0 and 2 are first coupled to the 
intermediate angular momentum $J_2$. This coupling scheme defines another 
orthonormal base in the whole space, 
\begin{eqnarray}
  \ket{B_{J_2};JM} \equiv \ket{j,(\frac{1}{2}j)J_2;JM}\ 
                      \ (J_2=j_+,j_-), 
                             \label{B_base}
\end{eqnarray}
where the three systems are ordered in the same way as in 
Eq.~(\ref{A_base}). The state $\ket{B_{J_2};JM}$ has the following exchange 
symmetries:  
\begin{eqnarray}
     \CS_{N+1}(02) \ket{B_{j_+};JM} &=& \ket{B_{j_+};JM}, 
                              \nonumber \\
     \CS_{N+1}(02) \ket{B_{j_-};JM} &=& 0.
                       \label{B_base_symmetry}
\end{eqnarray}

For a given set of $J(\ne 2j+1/2)$ and $M$, the two bases (\ref{A_base}) and 
(\ref{B_base}) are related by a unitary matrix, which can be taken 
to be real by the standard phase convention, 
\begin{eqnarray}
    \ket{A_{J_1};JM} = \sum_{J_2=j_+,j_-} R^J_{J_1J_2}\, \ket{B_{J_2};JM},
\end{eqnarray}
where the recoupling coefficient $R^J_{J_1J_2}$ is expressed by the 
Racah coefficient, 
\begin{eqnarray}
    R^J_{J_1J_2}= \sqrt{(2J_1+1)(2J_2+1)}\,W(j\frac{1}{2}Jj;J_1J_2),
\end{eqnarray}
and its explicit form is given by the following $2\times 2$ orthogonal matrix:
\begin{eqnarray}
    R^J = \left( 
       \begin{array}{cc}
           \frac{J+\frac{1}{2}}{2j+1} & 
                \frac{\sqrt{(2j+J+\frac{3}{2})(2j-J+\frac{1}{2})}}{2j+1} \\
           \frac{\sqrt{(2j+J+\frac{3}{2})(2j-J+\frac{1}{2})}}{2j+1} &
               -\frac{J+\frac{1}{2}}{2j+1} \\
       \end{array} \right), \nonumber \\
       { } \label{racah}
\end{eqnarray}
where rows and columns are allocated in the descending order of 
$J_1$ and $J_2$, respectively \cite{Rose57}.

Now the no-error conditions (\ref{unambiguity_symmetric}) imply that 
$E_1$ is an operator in the space spanned by $\ket{B_{j_-};JM}$, which 
is annihilated by $\CS_{N+1}(02)$ as shown in Eq.~(\ref{B_base_symmetry}). 
Furthermore, $E_1$ can be assumed to be a $U(2)$ scalar owing to the 
argument in the preceding section. This means that $E_1$ is diagonal with 
respect to $J$ and is proportional to the identity for $M$.
Combining these two properties, we find that $E_1$ should have the form
\begin{eqnarray}
   E_1 = \sum_{J=\frac{1}{2}}^{2j-\frac{1}{2}} e_J
         \sum_{M=-J}^{J}  \ket{B_{j_-};JM}\bra{B_{j_-};JM},
\end{eqnarray}
where coefficients $e_J$ should be non-negative by the positivity of $E_1$.

An upper bound is further imposed on the coefficient $e_J$ by
the remaining condition $1 \ge E_1+TE_1T$ in 
Eq.(\ref{positivity_symmetric}). 
This condition can be written as
\begin{eqnarray}
   1 &\ge& \sum_{J=\frac{1}{2}}^{2j-\frac{1}{2}} e_J 
           \sum_{M=-J}^{J}  \Big(
             \ket{B_{j_-};JM}\bra{B_{j_-};JM} \nonumber \\
     &   & \hspace{70pt}
            +\ket{A_{j_-};JM}\bra{A_{j_-};JM}
                          \Big)
                     \nonumber \\
     & = & \sum_{J=\frac{1}{2}}^{2j-\frac{1}{2}} e_J 
           \sum_{M=-J}^{J}
           \sum_{J_1,J_2} \ket{B_{J_1};JM} O_{J_1J_2}^{(J)} \bra{B_{J_2};JM},
                     \nonumber \\
     &   & 
\end{eqnarray}
where the matrix $O_{J_1J_2}^{(J)}$ is expressed in terms of the recoupling 
coefficients as follows:
\begin{eqnarray}
   O^{(J)} = \left(
       \begin{array}{cc}
           \left(R^J_{j_-j_+}\right)^2  &
                                R^J_{j_-j_+}R^J_{j_-j_-}   \\
            R^J_{j_-j_+}R^J_{j_-j_-}      &
                             1 + \left(R^J_{j_-j_-}\right)^2 \\
       \end{array}
       \right).
\end{eqnarray}
Eigenvalues of $O^{(J)}$ are readily calculated and found to be 
$1 \pm |R^J_{j_-j_-}|$. Thus constraints on the coefficient $e_J$ are 
given by
\begin{eqnarray}
             0 \le e_J \le \frac{1}{1+|R^J_{j_-j_-}|}.
                               \label{eJ_constraints}
\end{eqnarray}

Now it is easy to express the trace in Eq.(\ref{pS_symmetric}) in terms of 
the recoupling coefficients and $e_J$:
\begin{eqnarray}
 & &  \hspace{-30pt}
  \tr{E_1\CS_{N+1}(01)\CS_N(2)} 
                      \nonumber \\
 &=&  \sum_{JM} \bra{A_{j_+};JM}E_1\ket{A_{j_+};JM}
                      \nonumber \\
 &=&  \sum_{JM} \left( R^J_{j_+j_-} \right)^2 
                   \bra{B_{j_-};JM}E_1\ket{B_{j_-};JM}
                      \nonumber \\
 &=& \sum_{J=\frac{1}{2}}^{2j-\frac{1}{2}} 
             (2J+1)\left( R^J_{j_+j_-} \right)^2 e_J.
\end{eqnarray}
Therefore the probability, Eq.(\ref{pS_symmetric}), reaches its maximum 
when the coefficients $e_J$ takes its upper bound given in 
Eq.(\ref{eJ_constraints}).

Thus the optimal mean unambiguous identification probability is given by 
\begin{eqnarray}
   p_{\max}^{(N)}(d=2) &=& \frac{2J+1}{2_{N+1}2_N}
         \sum_{J=\frac{1}{2}}^{2j-\frac{1}{2}} 
             \frac{\left( R^J_{j_+j_-} \right)^2}
                         {1+|R^J_{j_-j_-}|}
           \nonumber \\
      &=& \frac{2J+1}{2_{N+1}2_N}\sum_{J=\frac{1}{2}}^{2j-\frac{1}{2}}
            \Big(1-|R^J_{j_-j_-}|\Big).
\end{eqnarray} 
We used the orthogonality of the recoupling matrix in the above derivation.
Inserting the explicit form of the recoupling coefficients and performing 
the sum in the above expression, we find a simple formula for 
$p_{\max}^{(N)}(d=2)$:
\begin{eqnarray}
  p_{\max}^{(N)}(d=2) = \frac{N}{3(N+1)}.
                \label{pmax2}
\end{eqnarray}
The optimal POVM is then given by
\begin{eqnarray}
         E_1 &=& \sum_{J=\frac{1}{2}}^{2j-\frac{1}{2}}
                  \frac{1}{1+|R^J_{j_-j_-}|}
         \sum_{M=-J}^{J}  \ket{B_{j_-};JM}\bra{B_{j_-};JM},
                      \nonumber \\
         E_2 &=& TE_1T,\ \ E_0=1-E_1-E_2.
\end{eqnarray}

As $N$ goes to infinity, $p_{\max}^{(N)}(d=2)$ approaches $1/3$, 
which is equal to the mean unambiguous discrimination probability 
$p_{\max}(d=2)$ given in Eq.~(\ref{p_discrimination}).
When $N=1$, on the other hand, the optimal POVM takes the form:
\begin{eqnarray}
 &&  E_1 = \frac{2}{3}( 1-\CS_2(02) ),\ E_2 = \frac{2}{3}( 1-\CS_2(01) ),
                      \nonumber \\ 
 &&  E_0 = 1-E_1-E_2.
\end{eqnarray}
which reproduces the one given by Bergou and Hillery \cite{Bergou05} in the 
case of the equal prior probabilities. For a two-spin-$1/2$-particle system, 
the state is either symmetric (triplet state) or antisymmetric (singlet 
state). Therefore, the optimal POVM for $N=1$ can also be written 
as
\begin{eqnarray}
    E_1 = \frac{2}{3}\ket{\Phi(02)}\bra{\Phi(02)},\ 
    E_2 = \frac{2}{3}\ket{\Phi(01)}\bra{\Phi(01)},
\end{eqnarray}
where we introduced the singlet state 
$\ket{\Phi(01)}=(\ket{0}\ket{1}-\ket{1}\ket{0})/\sqrt{2}$ for systems 
0 and 1 and similarly $\ket{\Phi(02)}$ for systems 0 and 2. 

One might wonder if we really need the complicated explicit form (\ref{racah}) 
of the Racah coefficients to obtain the simple final result of 
Eq.~(\ref{pmax2}). Actually we can avoid the explicit use of Racah 
coefficients if we exploit the algebraic properties of angular momentum 
operators. We will show it in the general dimensional case treated in the 
next section. 

\section{Case of arbitrary dimension $d$}
In this section, we generalize the argument in the preceding section to 
the arbitrary dimensional case. The essential point was the intimate 
relation between the symmetry properties under system permutations and 
the angular momentum of the combined system. The symmetry under system 
permutations is characterized by the representation of the symmetric 
group $S_{2N+1}$. And the angular momentum specifies the representation 
of $SU(2)$, more generally the unitary group $U(2)$. 
Therefore, in the case of arbitrary dimension $d$, we should classify 
the states according to representations of the symmetric group $S_{2N+1}$ 
and the unitary group $U(d)$.

Let us introduce the orthonormal base of the total space 
$(\BC^d)^{\otimes (2N+1)}$ according to irreducible representations 
of the symmetric group $S_{2N+1}$ and the unitary group $U(d)$.
We write states in this base as
\begin{eqnarray}
     \ket{\lambda,a,b}.
\end{eqnarray}
Here $\lambda$ represents an irreducible representation 
of $S_{2N+1}$, which is specified by a Young diagram. 
By the expression $\lambda=[\lambda_1,\lambda_2,\ldots]$, we denote a 
Young diagram consisting of a set of rows with their lengths given 
by $\lambda_1,\lambda_2,\ldots$. 
The label $a$ indexes orthogonal vectors in a particular $S_{2N+1}$ 
representation space 
and it runs from 1 to the dimension of the $S_{2N+1}$ representation.
It is known that the $\lambda$ also specifies irreducible representations 
of the unitary group $U(d)$ and its vectors are indexed by $b$, 
which runs from 1 to $m_\lambda(d)$, 
the multiplicity of representation $\lambda$ of $S_{2N+1}$ on 
$(\BC^d)^{\otimes (2N+1)}$ \cite{Hamermesh62}.  

\begin{figure}
\includegraphics[width=8cm]{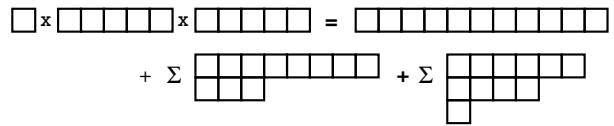}
\caption{\label{figyoung}
Decomposition of the product of three $U(d)$ 
irreducible representations $[1]\otimes[N]\otimes[N]$. 
The decomposition leads to the three orthogonal subspaces 
$V_n\ (n=1,2,3)$ according to the number of rows, $n$, of the Young's diagram. 
}
\end{figure}

As stated before, we can work in the subspace $V_{\rm sym}$, 
where systems 1 and 2 
are both totally symmetric, $\CS_{N}(1)=1$ and $\CS_{N}(2)=1$.
Possible Young diagrams $\lambda$ appearing in $V_{\rm sym}$ 
and the range of the index $a$ associated with a particular $\lambda$ 
can be determined by decomposing the product of three $U(d)$ 
irreducible representations $[1]\otimes[N]\otimes[N]$.
We decompose the space $V_{\rm sym}$ into three orthogonal subspaces 
$V_n\ (n=1,2,3)$ according to the number of rows, $n$, of the Young's diagram 
(see Fig. \ref{figyoung}). 

The subspace $V_1$ consists of totally symmetric states:
\begin{eqnarray}
    \ket{[2N+1],b},\ \ b=1,\ldots,m_{[2N+1]}(d),
\end{eqnarray}
where we omitted the index $a$, since the totally symmetric representation 
of $S_{2N+1}$ is one dimensional.
The states in $V_2$ belong to representations specified by 
Young's diagrams of two rows $[\lambda_1,\lambda_2]$, where 
$N+1 \le \lambda_1 \le 2N$ and $\lambda_2=2N+1-\lambda_1$. 
Since each of these $U(d)$ representations appears twice in $V_{\rm sym}$, 
we distinguish the two by label $a=1,2$ as follows:  
\begin{eqnarray}
   \ket{[\lambda_1,\lambda_2],a,b},\ \ 
    a=1,2,\ \  
    b=1,\ldots,m_{[\lambda_1,\lambda_2]}(d).
\end{eqnarray}
The remaining states are those whose Young's diagram has three rows and 
span the subspace $V_3$. 
\begin{eqnarray}
   \ket{[\lambda_1,\lambda_2,1],b},\ \ 
   b=1,\ldots,m_{[\lambda_1,\lambda_2,1]}(d),
\end{eqnarray}
where $N \le \lambda_1 \le 2N-1$ and $\lambda_2=2N-\lambda_1$.
We do not need the label $a$ for these states, because each representation 
of this type occurs only once in $V_{\rm sym}$. Note that the length of the 
third row is always 1.

Now let us determine a possible form of the POVM elements $E_1$ and $E_2$. 
First of all, $E_1$ should respect the no-error conditions, 
Eq.~(\ref{unambiguity_symmetric}). 
If $\ket{x}$ is in $V_1$, it is clear that $\CS_{N+1}(02)\ket{x}=\ket{x}$.
It is also easy to see that $\CS_{N+1}(02)\ket{x}=0$ for $\ket{x} \in V_3$, 
because representations with the Young's diagram of three rows cannot be 
constructed otherwise. 
States in $V_2$ for a given set of $\lambda$ and $b$ can be constructed 
in two different ways. We can assume the label $a=1,2$ for states in $V_2$ 
is chosen such that 
\begin{eqnarray}
    \CS_{N+1}(02) \ket{[\lambda_1,\lambda_2],1,b} &=& 0, 
                           \nonumber \\
    \CS_{N+1}(02) \ket{[\lambda_1,\lambda_2],2,b} &=& 
    \ket{[\lambda_1,\lambda_2],2,b}. 
\end{eqnarray}
We should also remember that the POVM element $E_1$ can be chosen to be a 
scalar with respect to $U(d)$. All these facts lead to the following form 
for $E_1$:
\begin{eqnarray}
    && E_1 = \sum_{\lambda_1} e_{[\lambda_1,\lambda_2]}
           \sum_b
           \ket{[\lambda_1,\lambda_2],1,b}\bra{[\lambda_1,\lambda_2],1,b}
                        \nonumber \\
    &+& \sum_{\lambda_1} e_{[\lambda_1,\lambda_2,1]}
           \sum_b
           \ket{[\lambda_1,\lambda_2,1],b}\bra{[\lambda_1,\lambda_2,1],b},
                        \label{E1_sum}
\end{eqnarray}
where $e_\lambda$ are some non-negative coefficients. 
The coefficient $e_\lambda$ depends only on $\lambda$, and the orthonormal 
states $\ket{[\lambda_1,\lambda_2],1,b}$ and $\ket{[\lambda_1,\lambda_2,1],b}$ 
are complete in the space annihilated by the projector $\CS_{N+1}(02)$. 
Therefore, this $E_1$ can be most conveniently expressed as the following 
operator form:
\begin{eqnarray}
    E_1 &=& e \Big( 1-\CS_{N+1}(02) \Big), \nonumber \\
      e &=& \sum_\lambda e_\lambda \Gamma_\lambda, 
                        \label{E1_operator}
\end{eqnarray}
where $\Gamma_\lambda$ is the projection operator onto the $U(d)$ 
representation space specified by $\lambda$. 
Furthermore, we can express $E_2$ as
\begin{eqnarray}
   E_2 = e \Big( 1-\CS_{N+1}(01) \Big), \label{E2_operator}
\end{eqnarray}
by the same operator $e$, since $\Gamma_\lambda$ is symmetric 
under the exchange of systems 1 and 2 and we can assume $E_2=TE_1T$ owing  
to the conclusion in Sec.~II.

Now that we have determined the possible form of $E_1$ and $E_2$, we 
can proceed to the positivity condition of $E_0$: namely, $1 \ge E_1 + E_2$. 
This condition can be written as 
\begin{eqnarray}
    1 \ge E_1+E_2 &=& e(2-\CS_{N+1}(01)-\CS_{N+1}(02))
                          \nonumber \\
                  &=& e(1-A),
                     \label{E0_operator}
\end{eqnarray}
where we introduced an operator $A$ in the subspace $V_{\rm sym}$ to be 
\begin{eqnarray}
     A \equiv \CS_{N+1}(01)+\CS_{N+1}(02)-1.
\end{eqnarray}

It is convenient to introduce another operator, which is the difference of 
the two projectors:
\begin{eqnarray}
    D &\equiv& \CS_{N+1}(01) - \CS_{N+1}(02).
\end{eqnarray}
Note that operators $A$ and $D$ are diagonal with respect to $\lambda$ and 
proportional to identity for the index $b$, since these operators involve  
just permutation operators. We also observe the relations
\begin{eqnarray}
    && A^2 = 1 - D^2, \\
    && AD+DA = 0,
\end{eqnarray}
which can be shown by an explicit calculation using 
$\CS_{N+1}(01)^2=\CS_{N+1}(01)$ and $\CS_{N+1}(02)^2=\CS_{N+1}(02)$.

The operator $A$ is $-1$ in the subspace $V_3$, since both projectors 
$\CS_{N+1}(01)$ and $\CS_{N+1}(02)$ annihilate any states in $V_3$. 
In the subspace $V_2$, two eigenvalues of $A$ have opposite signs 
in the invariant subspace associated with a given set of $\lambda$ and $b$. 
This is because $A$ and $D$ anticommute and the operation of $D$ changes 
the sign of eigenvalue of $A$. Note that $D$ does not annihilate any state 
in $V_2$. Combining these facts, we conclude that the positivity condition 
of (\ref{E0_operator}) implies the following inequality:
\begin{eqnarray}
      \frac{1}{1+|A|} \ge e,
                 \label{e_condition}
\end{eqnarray}
in subspaces $V_2$ and $V_3$.

Let us go back to the mean success probability, Eq.(\ref{pS_symmetric}), 
and write it with the form of $E_1$ given by Eq.(\ref{E1_operator}).
\begin{eqnarray}
 p^{(N)}(d) = \frac{1}{d_{N+1}d_N}
                \tr{e(1-\CS_{N+1}(02))\CS_{N+1}(01))},
\end{eqnarray}
which can be further rewritten as 
\begin{eqnarray}
 p^{(N)}(d) = \frac{1}{2d_{N+1}d_N}
                \tr{e(1-A^2)}.
\end{eqnarray}

In the above equation, we find that the subspaces $V_1$ and $V_2$  have no 
contribution to the trace sum. 
And $|A|$ in the upper bound of $e$ in Eq.(\ref{e_condition}) commutes 
with $1-A^2$ in the trace.  Therefore, we immediately obtain the optimal 
mean success probability as follows:
\begin{eqnarray}
   p^{(N)}(d) &\le& \frac{1}{2d_{N+1}d_N}
             \tr{\frac{1}{1+|A|}(1-A^2)}
                             \nonumber \\
              &=& \frac{1}{2d_{N+1}d_N}\tr{1-|A|}
                             \nonumber \\
              &\equiv& p^{(N)}_{\max}(d).
\end{eqnarray}
The optimal success probability is thus attained by 
\begin{eqnarray}
   E_1 &=& \frac{1}{1+|A|}(1-\CS_{N+1}(02)), 
                               \nonumber \\ 
   E_2 &=& \frac{1}{1+|A|}(1-\CS_{N+1}(01)), 
                               \nonumber \\
   E_0 &=& \frac{A+|A|}{1+|A|}.
\end{eqnarray} 
Here we took $e_{[\lambda_1,\lambda_2,1]} = \frac{1}{2}$ for simplicity, 
which is the maximum value allowed by Eq.(\ref{e_condition}), 
though the subspace $V_3$ does not contribute to $p^{(N)}_{\max}(d)$.

We must still determine eigenvalues of $|A|$ in the subspace $V_2$ in order to 
evaluate $p^{(N)}_{\max}(d)$ further. As mentioned above, the operator $A$ is 
proportional to identity with respect to the index $b$ for a given particular 
$\lambda$. Therefore, eigenvalues of $A$ are independent of the dimension $d$ 
up to multiplicity. Thus we can assume the dimension $d$ is equal to 2, 
which allows us to exploit the angular momentum algebra. 

Assuming $d=2$, we introduce three sets of angular momentum operators: 
$\bold{s}(0) \equiv \frac{1}{2}\bold{\sigma}(0)$ for system 0, 
$\bold{j}(1)$ for system 1, and $\bold{j}(2)$ for system 2. 
In the subspace $V_{\rm sym}$, we have 
$\bold{j}(a)^2=\frac{N}{2}(\frac{N}{2}+1)$, 
since the total angular momentum of system $a=1,2$ is $\frac{N}{2}$. 
It is then easy to show that the projector $S_{N+1}(0a)$ can be 
written in terms of the angular momentum operators as follows ($a=1,2$):
\begin{eqnarray}
  S_{N+1}(0a) = \frac{1}{N+1} 
        \left( 2\bold{j}(a) \cdot \bold{s}(0) + \frac{N}{2} + 1 \right). 
\end{eqnarray}
Using this form for the projectors, we calculate $A^2$. 
After some algebraic calculation involving the Pauli matrices and 
angular momentum commutation relations, we find  
\begin{eqnarray}
 A^2 = \frac{1}{(N+1)^2} \left( \bold{J}^2 + \frac{1}{4} \right),
\end{eqnarray}
where $\bold{J} = \bold{s}(0)+\bold{j}(1)+\bold{j}(1)$ is the total 
angular momentum operator. The eigenvalue of $\bold{J}^2$ is 
$J(J+1)\ (J=\frac{1}{2},\ldots,N+\frac{1}{2})$, which implies that eigenvalues 
of $A$ are given by $\pm\frac{J+\frac{1}{2}}{N+1}$ with multiplicity $2J+1$.

For a general dimension $d$, we thus conclude that $A$ in $V_2$ has 
eigenvalues $\pm\frac{\lambda_1-N}{N+1}$ with multiplicity 
$m_{[\lambda_1,\lambda_2]}(d)$, 
since the total angular momentum $J$ in the case of $d=2$ is given by 
$J=\frac{\lambda_1-\lambda_2}{2}= \lambda_1-N-\frac{1}{2}$.

Finally we obtain the formula for the optimal success probability 
\begin{eqnarray}
  p^{(N)}_{\max}(d) &=& \frac{1}{d_{N+1}d_N}\times  \nonumber \\
      && \sum_{\lambda_1=N+1}^{2N} 
         m_{[\lambda_1 \lambda_2]}(d) 
         \left( 1 - \frac{\lambda_1-N}{N+1} \right),
                     \label{pNd_final}
\end{eqnarray}
where $m_{[\lambda_1,\lambda_2]}(d)$ ($\lambda_2=2N+1-\lambda_1$) is 
the multiplicity of the $S_{2N+1}$ irreducible 
representation $[\lambda_1,\lambda_2]$ and given by \cite{Hamermesh62}
\begin{eqnarray}
  m_{[\lambda_1,\lambda_2]}(d) = \frac{(\lambda_1+d-1)!(\lambda_2+d-2)!
                                       (\lambda_1-\lambda_2+1)}
                          {(d-1)!(d-2)!(\lambda_1+1)!\lambda_2!}.
             \nonumber \\
\end{eqnarray}

\begin{figure}
\includegraphics[width=8cm]{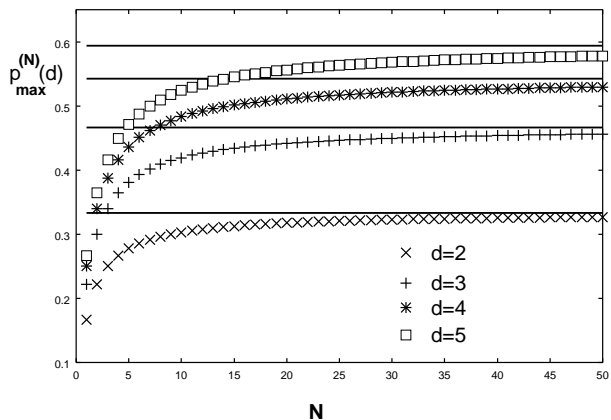}
\caption{\label{unfig}
The optimal mean unambiguous identification probability 
$p^{(N)}_{\max}(d)$ as a function of the number of the copies ($N$) of the 
reference states.
As $N$ increases,  $p^{(N)}_{\max}(d)$ approaches 
the mean optimal unambiguous discrimination probability shown by the horizontal  lines. 
}
\end{figure}

Let us study the asymptotic value of $p^{(N)}_{\max}(d)$ when the number 
of the copies $N$ is very large. In this case we can replace the sum in 
Eq.(\ref{pNd_final}) by a continuous integration with respect to 
$x = \frac{\lambda_1}{N}-1$. We find 
\begin{eqnarray}
  p^{(N)}_{\max}(d) &\rightarrow& 
             2(d-1)\int_0^1\!dx\, (1+x)^{d-2}(1-x)^{d-1}
                            \nonumber \\
                    &=& 
             1 - \frac{2^{d-1}(d-1)!}{(2d-1)!!}\ \ 
                             (N \rightarrow \infty),
\end{eqnarray}
which is equal to $p_{\max}(d)$ given by Eq.~(\ref{p_discrimination}). 
Thus, as expected, the unambiguous identification 
reduces to the unambiguous discrimination as the number of the copies goes 
to infinity. Figure \ref{unfig} displays how the unambiguous identification 
probability approaches the unambiguous discrimination probability as the 
number of the copies increases.

\section{Concluding remarks}
We have studied the problem of unambiguously identifying the input state of a 
$d$-dimensional system with one of the two reference states 
when $N$ copies of each reference state are presented with no classical 
information. 
We have determined the optimal mean unambiguous identification probability 
$p_{\max}^{(N)}(d)$ as a function of $d$ and $N$.

It is interesting to compare the results in this paper and those for the 
identification problem without the no-error conditions, which was studied 
in our previous paper \cite{Hayashi05_identification}. 
In both problems the symmetry under system permutations plays an essential 
role. 
This is also true in the state comparison studied by Barnett, 
Chefles, and Jex \cite{Barnett03}, in which one's task is to establish 
whether or not two quantum systems have been prepared in the same state. 
In this case the symmetry under exchanging the two systems can characterize 
the optimal POVM. In the state identification, however, we must distinguish 
the two states 
$\rho_1(0) \rho_1^{\otimes N}(1) \rho_2^{\otimes N}(2)$ and 
$\rho_2(0) \rho_1^{\otimes N}(1) \rho_2^{\otimes N}(2)$. 
Therefore, we must consider the symmetries with respect to partial 
permutations among systems 0 and 1 and among systems 0 and 2. 
The relevant operators are noncommutable projection operators $\CS_{N+1}(01)$ 
and $\CS_{N+1}(02)$, which makes the optimization of the success probability 
rather involved. The success probability is expressed by the trace of the 
modulus of some linear combination of the symmetrizers $\CS_{N+1}(01)$ and 
$\CS_{N+1}(02)$: $D=\CS_{N+1}(01)-\CS_{N+1}(02)$ in the case of the 
identification problem without the no-error conditions and 
$A=\CS_{N+1}(01)+\CS_{N+1}(02)-1$ in the unambiguous identification problem 
considered in this paper. 

As for the optimal POVM, it was shown that the optimal success probability 
can be attained by a projective measurement in the identification problem 
without the no-error conditions, whereas the optimal POVM obtained for the 
unambiguous identification considered here is not a projective 
measurement. 


\end{document}